\pdfoutput=1
\documentclass[11pt,aps,a4paper,eqsecnum,amsmath,amssymb,longbibliography,notitlepage,nofootinbib]{revtex4-1}

\usepackage{color}
\usepackage{graphicx}
\usepackage{hyperref}

\usepackage{amsmath}
\usepackage{bm}

\usepackage{verbatim}
\usepackage[dvipsnames]{xcolor}
\newcommand{\bra}[1]{\langle #1|}
\newcommand{\ket}[1]{|#1\rangle}
\newcommand{\braket}[2]{\left\langle #1|#2\right\rangle}

\usepackage{tikz}

\makeatletter
\def\@bibdataout@aps{%
\immediate\write\@bibdataout{%
@CONTROL{%
apsrev41Control%
\longbibliography@sw{%
    ,author="08",editor="1",pages="1",title="0",year="1"%
    }{%
    ,author="08",editor="1",pages="1",title="",year="1"%
    }%
  }%
}%
\if@filesw \immediate \write \@auxout {\string \citation {apsrev41Control}}\fi 
}
\makeatother

\begin{document}
\preprint{}
\title{
  Factorization of density matrices in the critical RSOS models
}  
\author{Daniel Westerfeld}
\author{Maxime Großpietsch}
\author{Hannes Kakuschke}
\author{Holger Frahm}
\affiliation{%
Institut f\"ur Theoretische Physik, Leibniz Universit\"at Hannover,
Appelstra\ss{}e 2, 30167 Hannover, Germany}

\date{\today}

\begin{abstract}
We study reduced density matrices of the integrable critical RSOS
model in a particular topological sector containing the ground state.
Similar as in the spin-$1/2$ Heisenberg model it has been observed that correlation functions of this model on
short segments can be `factorized': they are completely determined by
a single nearest-neighbour two-point function $\omega$ and a set of
structure functions.  While $\omega$ captures the
dependence on the system size and the state of the system the structure functions can be expressed in terms of the
possible operators on the segment, in the present case representations
of the Temperley-Lieb algebra $\text{TL}_n$, and are independent of the model
parameters.  We present explicit results for the function $\omega$ in
the infinite system ground state of the model and compute multi-point
local height probabilities for up to four adjacent sites for the RSOS
model and the related three-point correlation functions of non-Abelian
$su(2)_k$ anyons.
\end{abstract}


\maketitle

\section{Introduction}
The essential step from a theoretical model for a quantum many-body system to the description of experimental observations is the computation of correlation functions.  Already the information about the energy levels of a system with two-body interactions is encoded in the two-particle reduced density matrix (RDM) in a given eigenstate. Similarly, knowing the RDMs for a few particles in a given $N$-particle state gives access to many properties of this system.  In general, however, correlations due to interactions and the statistics of the constituents of the system lead to restrictions on RDMs which pose a challenge to perturbative approaches for their calculation \cite{RDM_book}.

On the other hand, for certain integrable models constructed from an $R$-matrix satisfying a Yang-Baxter  equation -- in particular the six-vertex model and the related spin-$1/2$ chains -- a growing number of exact results for correlation functions and RDMs has been obtained by making use of the underlying mathematical structures \cite{JMMN92,JiMi96,KiMT00,GoKS04}: multiple integral formulations following from the representation of vertex operators realizing quantum affine symmetries or from the algebraic Bethe ansatz and functional equations of $q$-Knizhnik-Zamolodchikov (qKZ) type provide explicit expressions. Their efficient evaluation using numerical methods, however, remained to be an obstacle. This situation has improved significantly when it was shown that the multiple integral representations of density matrices on short segments can be factorized into single ones \cite{BGKS06} and that $N$-point correlation functions (as well as RDMs) of an inhomogeneous generalization of the isotropic Heisenberg spin chain can be written in terms of a nearest neighbour two-point function $\omega$ (the `physical part') and a set of recursively defined `structure functions' (or `algebraic part') $f_{N;I,J}$ of the spectral parameters $\lambda_j$ \cite{BJMST06a}:
\begin{equation}
\label{XXXfac}
 D_N(\lambda_1,\dots,\lambda_N) = \sum_{m=0}^{[N/2]} \sum_{I,J} \left(\prod_{p=1}^m \omega(\lambda_{i_p},\lambda_{j_p}) \right) f_{N;I,J}(\lambda_1,\dots,\lambda_N)
\end{equation}
where $I=(i_1,\dots,i_m)$ and $J=(j_1,\dots,j_m)$ such that $I \cap J=\varnothing$, $1\leq i_p<j_p\leq N$ and $i_1<\dots <i_m$. 
Similar expressions have been proven for a general inhomogeneous six-vertex model (including the finite temperature and the finite length Heisenberg chain as special cases) using the fermionic structure in the space of operators of this model \cite{Boos07_HiddenGrassmann,Boos09a_HiddenGrassmann_ii}.  Based on (\ref{XXXfac}) it has been argued that correlation functions in excited states of the Heisenberg model factorize if the physical part is changed appropriately \cite{Pozsgay17}.

With Eq.~(\ref{XXXfac}) and its extension using the fermionic basis approach there exists a powerful tool to compute correlation functions in integrable vertex models, see e.g.\ \cite{Sato.etal11,GoKW21}.  Another class of Yang-Baxter integrable models, so-called interaction-round-a-face (or face) models, has attracted considerable interest recently as such models can be used to describe the collective behaviour of interacting non-Abelian anyons in topological quantum liquids, see e.g.\ \cite{FTLT07,GATL09,GATH13,FiFr13,Finch.etal14}.  For such models the development of a similar framework for the computation of RDMs is only at its beginning: for the restricted solid-on-solid (RSOS) models one-point functions such as local height probabilities (LHPs) have been computed in the thermodynamic limit using Baxter's corner transfer matrix \cite{AnBF84}. Moreover, qKZ equations for correlation functions of vertex operators related to quantum group symmetries and multiple integral representations for multi-point LHPs have been  constructed in the massive phases of the  RSOS models \cite{FJMM94,LuPu96}. Further results have been obtained for face models with a dynamical $R$-matrix allowing for the transfer of concepts such as the algebraic Bethe ansatz or separation of variables from vertex models \cite{FeVa96a,LeTe14,LeNT16a}.

Recently, we have expressed reduced density matrices in general face models in terms of their local Boltzmann weights \cite{FrWe21}.  Following a similar construction for the Heisenberg spin chain \cite{AuKl12} this allows to derive discrete functional equations called `reduced' qKZ equations satisfied by the RDMs.  A study of these equations for the critical RSOS models has been initiated in \cite{FrWe21}: based on finite size studies of these models with a small number of allowed local heights we found that their two- and three-site RDMs can be written in factorized form similar to (\ref{XXXfac}) for \emph{all} states from a particular topological sector. In this paper we continue this work: after a short review of the definition of the models we propose the functional equations satisfied by the RDMs of the RSOS models. Complementing the restricted qKZ equation with a set of recurrence and reduction relations resulting from the properties of the Boltzmann weights together with an identity for the asymptotics of the RDMs observed to hold in a particular topological sector allows to compute the algebraic part of the two- and three-site density matrices which then can be expressed in terms of the generators of the Temperley-Lieb algebra underlying the RSOS model.  For the physical part we solve the restricted qKZ equation for the nearest-neighbour two-site RDM for the ground state of the  model in the thermodynamic limit and compare our result to those for finite systems.  Finally, we apply our expressions to compute multi-point LHPs for the critical RSOS models.

\section{The critical RSOS models}
The RSOS models are defined on a square lattice where the spins (or heights) lying on the vertices take values from the set $\mathfrak{S}=\{1,2,\dots,r-1\}$ \cite{AnBF84}.  Spins $a$ and $b$ on neighbouring sites are constrained by the adjacency condition $|a-b|=1$. The Boltzmann weights for an elementary face where this constraint is satisfied on all four edges are defined as
\begin{equation}
\label{eq:rsos_weights}
  W \left(
    \begin{matrix}
      a & b\\
      c & d
    \end{matrix}\,\middle|\, 
    u\right)=\delta_{ad}\sqrt{\frac{g_b g_c}{g_a g_d}}\rho(u+\lambda)-\delta_{bc} \rho(u)
    =
  \begin{tikzpicture}[baseline=(current bounding box.center)]
    \draw (1,0.5)--(2,0.5);
    \draw (1,-0.5)--(2,-0.5);
    
    \draw (1,-0.5)--(1,0.5);
    \draw (2,-0.5)--(2,0.5);
    
    \node [above] at (1,0.5) {$a$} ;
    \node [below] at (1,-0.5) {$c$} ;
    \node [above] at (2,0.5) {$b$} ;
    \node [below] at (2,-0.5) {$d$} ;
    \node at (1.5,0) {$u$};
  \end{tikzpicture}=\begin{tikzpicture}[baseline=(current bounding box.center)]
  \def \d {0.5 * 1.41421356237}
  \draw (0,0)--(\d,\d)--(2*\d,0)--(\d,-\d)--(0,0);
  \node [left] at (0,0) {$c$};
  \node [above] at (\d,\d) {$a$};
  \node [right] at (2*\d,0) {$b$};
  \node [below] at (\d,-\d) {$d$};
  \node at (\d,0) {$u$};
\end{tikzpicture}\,
\end{equation}
with the crossing parameter $\lambda=\pi/r$ and
\begin{equation}\label{eq:gauge-factors}
  \rho(u)=\frac{\sin(u-\lambda)}{\sin\lambda},\quad g_x=\frac{\sin\lambda x}{\sin\lambda}\, .
\end{equation}
They satisfy the unitarity condition (dotted lines in the graphical notation indicate that the connected heights are taken to be equal, heights at nodes marked by a solid circle are summed over)
\begin{align}\label{eq:uniface}
  \sum_{e\in\mathfrak{S}} W \left(
    \begin{matrix}
      d & e\\
      a & b
    \end{matrix}\,\middle|\, 
    u\right)
    W \left(
    \begin{matrix}
      d & c\\
      e & b
    \end{matrix}\,\middle|\, 
    -u\right) = 
  \begin{tikzpicture}[baseline=(current bounding box.center)]
  \draw (0,0)--(0.5,0.5)--(1,0)--(1.5,0.5)--(2,0)--(1.5,-0.5)--(1,0)--(0.5,-0.5)--(0,0);
  \draw  [dotted] (0.5,0.5)--(1.5,0.5);
  \draw  [dotted] (0.5,-0.5)--(1.5,-0.5);
  \node[left] at (0,0) {$a$};
  \node[above] at (0.5,0.5) {$d$};
  \node[above] at (1.5,0.5) {$d$};
  \node[right] at (2,0) {$c$};
  \node[below] at (1.5,-0.5) {$b$};
  \node[below] at (1,0) {$e$};
  \node[below] at (0.5,-0.5) {$b$};
  \node at (0.5,0) {$u$};
  \node at (1.5,0) {$-u$};
  \node[draw,circle,inner sep=1pt,fill] at (1,0) {};
\end{tikzpicture}
  &=\rho(u)\rho(-u)\delta_{ac},
\end{align}
crossing symmetry
\begin{equation}\label{eq:crossface}
W \left(
    \begin{matrix}
      a & b\\
      c & d
    \end{matrix}\,\middle|\, 
    u\right) = \sqrt{\frac{g_b\,g_c}{g_a\,g_d}}\,W \left(
    \begin{matrix}
      b & d\\
      a & c
    \end{matrix}\,\middle|\, 
    \lambda-u\right)\,,
  \end{equation}
and the initial condition
\begin{equation}
\label{eq:initface}
 W \left(
    \begin{matrix}
      a & b\\
      c & d
    \end{matrix}\,\middle|\, 
    0\right)=\delta_{b,c}\,.
\end{equation}
With these local weights we define single row operators
\begin{equation}
  \label{eq:def_monodromy}
  \begin{aligned}
    \bra{\bm{a}} T{^{\alpha\beta}_{\gamma\delta}}(u) \ket{\bm{b}}
    &=
      \begin{tikzpicture}[baseline=(current bounding box.center),scale=0.9]
         \def \b {1.5};
         \draw (0,0+\b/2)--(0+4*\b,0+\b/2);
         \draw (0,0-\b/2)--(0+4*\b,0-\b/2);
         \foreach \x in {0,\b,3*\b,4*\b}{
         \draw (\x,\b/2)--(\x,-\b/2);}
         \node at (\b-0.5*\b,0) {$u-u_1$} ;
         \node at (4*\b-0.5*\b,0) {$u-u_L$} ;
         \node at (2*\b,0) {$\dots$} ;
         \node [above] at (0,\b/2) {${\alpha}=a_0$} ;
  \node [below] at (0,-\b/2) {${\gamma}=b_0$} ;
  \node [above] at (\b,\b/2) {$a_1$} ;
  \node [below] at (\b,-\b/2) {$b_1$} ;
  \node [above] at (3*\b,\b/2) {$a_{L-1}$} ;
  \node [below] at (3*\b,-\b/2) {$b_{L-1}$} ;
    \node [above] at (4*\b,\b/2) {$a_{L}={\beta}$} ;
  \node [below] at (4*\b,-\b/2) {$b_{L}={\delta}$} ;
     \end{tikzpicture}\,
  \end{aligned}
\end{equation}
acting on the space $\mathcal{H}^L=\text{span}\left\{\ket{a_0\dots a_L}:|a_{j+1}-a_j|=1\right\}$.  Imposing periodic boundary conditions in the horizontal direction and performing the trace over the spins $\alpha=\beta$ and $\gamma=\delta$ one obtains the transfer matrix of the inhomogeneous RSOS model
\begin{equation}
\label{eq:transfer}
    t(u) = \sum_{\alpha\gamma} T^{\alpha\alpha}_{\gamma\gamma}(u)\,.
\end{equation}
where the inhomogeneities $u_k\in\mathbb{C}$ parameterize local variations in the interactions of the model. 

The RSOS models are exactly solvable: the transfer matrix (\ref{eq:transfer}) commutes for different values of the spectral parameter $u$ as a consequence of the Boltzmann weights satisfying the Yang-Baxter equation
\begin{equation}\label{eq:ybeface}
\begin{aligned} 
    &\sum\limits_{g\in\mathfrak{S}}   W \left(
    \begin{matrix}
      f & g\\
      a & b
    \end{matrix}\,\middle|\, 
    u-v\right)   W \left(
    \begin{matrix}
      f & e\\
      g & d
    \end{matrix}\,\middle|\, 
    v\right)
      W \left(
    \begin{matrix}
      g & d\\
      b & c
    \end{matrix}\,\middle|\, 
    u\right)\\
    &\qquad = \sum\limits_{g\in\mathfrak{S}} W \left(
    \begin{matrix}
      f & e\\
      a & g
    \end{matrix}\,\middle|\, 
    u\right)
      W \left(
    \begin{matrix}
      a & g\\
      b & c
    \end{matrix}\,\middle|\, 
    v\right)
    W \left(
    \begin{matrix}
      e & d\\
      g & c
    \end{matrix}\,\middle|\, 
    u-v\right)\,.
    \end{aligned}
\end{equation}
Alternatively this equation can be expressed as
\begin{equation}
  W_j(u)\,W_{j+1}(u+v)\,W_j(v) = W_{j+1}(v)\,W_{j}(u+v)\,W_{j+1}(u)\,,
\end{equation}
in terms of the Yang-Baxter operators $W_j(u)$ acting on $\mathcal{H}^L$ as
\begin{equation}
\label{eq:YB-op}
        \bra{a_0\dots a_L}W_j(u)\ket{b_0\dots b_L} \equiv \prod_{k\ne j}\delta_{a_kb_k}\,W \left(
    \begin{matrix}
      a_{j-1} & a_j\\
      b_j & a_{j+1}
    \end{matrix}\,\middle|\, 
    u\right)\,.
\end{equation}
With  (\ref{eq:rsos_weights}) these operators can be expanded as
\begin{equation}
    W_j(u) = \rho(u)\,\mathbf{1} + \rho(u+\lambda)\,e_j,
\end{equation}
where $\{\mathbf{1},e_1,\dots,e_{n-1}\}$ is a representation of the generating elements of the Temperley-Lieb algebra $\text{TL}_n(\beta)$
\begin{equation}
    \label{eq:TL}
    \begin{aligned}
    &e_j^2 = \beta\,\, e_j\,, \quad e_j=e_j\,e_{j\pm1}\,e_j\,,\\
    &e_j e_{j'} = e_{j'} e_j \quad \text{for~}|j-j'|>1\,,
    \end{aligned}
\end{equation}
with $\beta=2\cos\lambda$. 

By construction the transfer matrix $t(u)$ and its eigenvalues $\Lambda(u)$ are Fourier polynomials of degree $L$
\begin{equation}\label{eq:fouriereigen}
 \Lambda(u)=\sum_{n=-L/2}^{L/2} \Lambda_{2n} \mathrm{e}^{i2nu}\,,
\end{equation}
where the leading Fourier coefficients are known to take values \cite{KlPe92}
\begin{equation}\label{eq:asymp_fouri}
   \Lambda_{\pm L}=\left(\prod_{k=1}^L \exp(\mp i (u_k+\lambda/2))\right)\frac{2\cos((2j+1)\lambda)}{(2\sin\lambda)^L}\,,
   \quad j\in\{0,\frac12,1,\dots,\frac{r-2}2\}\,.
\end{equation}
This allows to decompose the spectrum of the RSOS model into topological sectors with a `quantum dimension' labeled by the quantum number $j$
\begin{equation}
  d_q(j) = \frac{\sin(\pi(2j+1)/r)}{\sin(\pi/r)}\,.
\end{equation}
In view of applications we will be particularly interested in the RSOS model in the Hamiltonian limit: expanding the homogeneous transfer matrix, i.e.\ $u_k\equiv0$, around the shift point $u=0$ to first order one obtains the periodic Temperley-Lieb Hamiltonian of the one-dimensional quantum RSOS model \cite{BaRe89}:
\begin{equation}
\label{eq:HqRSOS}
    \mathcal{H} = \frac{\lambda}{4\pi\sin\lambda} \sum_{j=1}^L e_j\,.
\end{equation}
Note that this model has recently been used to study the collective excitations in a linear chain of interacting $su(2)_k$ non-Abelian anyons with spin-$1/2$ for $k=r-2$ \cite{FTLT07}.  Here the adjacency conditions for neighbouring heights are enforced by anyonic fusion rules.

In Ref.~\cite{FrWe21} we have shown that reduced density matrices in an eigenstate $\ket{\Phi}$ corresponding to the eigenvalue $\Lambda(u)$ of the transfer matrix (\ref{eq:transfer}) can be expressed in terms of the single row operators (\ref{eq:def_monodromy}): define local operators acting on sequences of adjacent sites $n_1,\dots,n_2$ through their matrix elements in the basis of $\mathcal{H}^L$ as
\begin{equation}
    \bra{\bm{a}}E^{\alpha_{n_1}\dots\alpha_{n_2}}_{\beta_{n_1}\dots\beta_{n_2}}\ket{\bm{b}} = \prod_{k=n_1}^{n_2} \delta_{a_k,\alpha_k}\;
    \delta_{b_k,\beta_k}\prod_{j\notin\{n_1\dots n_2\}}\delta_{a_j b_j}\,,
\end{equation}
and generalized RDMs $D_N$ depending on a set of auxiliary spectral parameters $\lambda_j$, $j=1,\dots,N$,
\begin{equation}
    \label{eq:def_D}
    D_N(\lambda_1,\dots,\lambda_{N})^{\{\underline{\bm{\alpha}}\}\{\underline{\bm{\beta}}\}}   
    = 
      \frac{\bra{\Phi} \prod_{k=1}^{N}  T^{\alpha_{k-1} \beta_{k-1}}_{\alpha_k \beta_k}(\lambda_k) \ket{\Phi}}
      {{\braket{\Phi}{\Phi}}\,\prod\limits_{k=1}^{N} \Lambda(\lambda_k)}\,,
\end{equation}
where $\underline{\bm{\alpha}} = (\alpha_0,\dots,\alpha_N)$ and $\underline{\bm{\beta}} = (\beta_0,\dots,\beta_N)$ are sequences of heights labeling the basis of the space $\mathcal{V}^N=\text{span}\{\ket{\alpha_0\dots\alpha_N}: |\alpha_{j+1}-\alpha_j|=1\}$ (for a graphical representation of this object see Appendix~\ref{app:DN_graphical}).
With these definitions one can show that
\begin{equation}
\label{eq:EtoD}
 \frac{1}{\braket{\Phi}{\Phi}}\,\bra{\Phi}\,E^{\alpha_{0}\dots\alpha_{N}}_{\beta_{0}\dots\beta_{N}}  \ket{\Phi}=D_N(\lambda_{1},\dots,\lambda_{N})^{\{\underline{\bm{\alpha}}\}\{\underline{\bm{\beta}}\}} \Big\vert_{\lambda_k=u_k,\,k=1,\dots,N}\,.
\end{equation}
In view of this relation we can complement (\ref{eq:def_D}) by
\begin{equation}
  \label{def_D0}
  D_0^{\alpha,\alpha} = \frac{1}{\braket{\Phi}{\Phi}}\,\bra{\Phi}\,E^{\alpha_{0}}_{\alpha_{0}}  \ket{\Phi} \equiv P_\alpha
\end{equation}
being the local height probability (LHP) of the critical RSOS model \cite{AnBF84}
\begin{equation}
\label{LHP1}
  P_\alpha = \frac{2\lambda}{\pi}\,\sin^2\alpha\lambda\,.
\end{equation}

As a consequence of the state $\ket{\Phi}$ satisfying periodic boundary conditions the matrix elements $D_N(\lambda_1,\dots,\lambda_{N})^{\{\underline{\bm{\alpha}}\}\{\underline{\bm{\beta}}\}}$ vanish for $\alpha_0\neq\beta_0$, $\alpha_N\neq\beta_N$. Therefore $D_N$ can be decomposed into blocks labeled $[\alpha_{0},\alpha_{N}]$, i.e.\
\begin{equation}
  \label{eq:def_Dblock}
  D_N(\lambda_1,\dots,\lambda_N)^{\{\underline{\bm{\alpha}}\}\{\underline{\bm{\beta}}\}}  = \left(D_N^{[\alpha_0,\alpha_N]}(\lambda_1,\dots,\lambda_N)\right)^{\alpha_1\dots\alpha_{N-1}}_{\beta_1\dots\beta_{N-1}}\,.
\end{equation}
Note that the same is true for representations of the Temperley-Lieb algebra $\text{TL}_N$ as operators on $\mathcal{V}^N$.

\section{Properties of the reduced density matrix}
\label{sec:funceq}
In \cite{FrWe21} a factorization of correlation functions similar to (\ref{XXXfac}) has been observed to hold for the generalized two- and three-site density matrices of RSOS models with $r=4$ and $5$ in the topological sectors with quantum dimension $d_q(j)=1$: 
\begin{equation}
\label{RSOSfac}
\begin{aligned}
    &D_1(\lambda_1) = F_{1;\varnothing,\varnothing}\,,\quad
    D_2(\lambda_1,\lambda_2) = F_{2;\varnothing,\varnothing} + F_{2;(1)(2)}\,\omega(\lambda_1,\lambda_2)\,,\\
    &D_3(\lambda_1,\lambda_2,\lambda_3) = F_{3;\varnothing,\varnothing} + F_{3;(1)(2)}(\lambda_1,\lambda_2,\lambda_3)\, \omega(\lambda_1,\lambda_2)\\
    &\qquad\qquad\qquad\qquad + F_{3;(2)(3)}(\lambda_1,\lambda_2,\lambda_3)\, \omega(\lambda_2,\lambda_3) + F_{3;(1)(3)}(\lambda_1,\lambda_2,\lambda_3)\, \omega(\lambda_1,\lambda_3)\,,
\end{aligned}    
\end{equation}
with a single symmetric two-point function $\omega(u,v)=\omega(v,u)$.  For states from these sectors the algebraic part, given in terms of the 'structure functions' $F_{N;I,J}(\lambda_1,\dots,\lambda_N)$, is independent of the specific model, i.e.\ of the system size and of the inhomogeneities $u_k$. The $F_{N;I,J}(\lambda_1,\dots,\lambda_N)$ are matrices acting on the space $\mathcal{V}^N$. Because of this particularly simple form of the RDMs we restrict ourselves to consider states from these sectors only.  Note that this includes the ground state of the RSOS model in the thermodynamic limit.  

Among the elements of the single-site density matrix $D_1(\lambda_1)$ of the critical RSOS models only the diagonal ones are non-zero (the $D_1$-blocks $[a,a+1]$ allowed by the adjacency rules are one-dimensional).
Moreover, $D_1(\lambda_1)$ has been shown to be independent of the spectral parameter $\lambda_1$ and can be obtained from the LHP (\ref{LHP1}), see Ref.~\cite{FrWe21}: using the symmetry $P_{a,a+1}=P_{a+1,a}$ of the two-site LHPs (i.e.\ the probabilities that the heights at two adjacent sites are equal to $a$ and $a+1$) and the identity $\sum_b P_{a,b}=P_a$ the non-zero elements of $D_1=F_{1;\varnothing,\varnothing}$ are found to  be ($1\leq a \leq r-2$)
\begin{equation}
\label{D1}
    P_{a,a+1}=\langle a,a+1| D_1(\lambda_1) | a,a+1\rangle
    = \langle a+1,a| D_1(\lambda_1) | a+1,a\rangle
    = \frac{\lambda \sin a \lambda\, \sin((a+1)\lambda)}{\pi\,\cos\lambda}\,.
\end{equation}

%

In  the following we collect several properties of the reduced density matrices constructed in the previous section and relations satisfied by  them which will be used below for their computation:

\paragraph{} By construction (\ref{eq:def_D}) $\Lambda(\lambda_j)\,D_N(\lambda_1,\dots,\lambda_N)$ is a Fourier polynomial of the spectral parameter $\lambda_j$.  Hence, the poles of $D_N$ correspond to zeroes of the transfer matrix eigenvalues $\Lambda(u)$.

\paragraph{} As a consequence of the YBE (\ref{eq:ybeface}) the arguments of $D_N(\lambda_1,\dots\lambda_N)$ can be reordered by application of the Yang-Baxter operators:
\begin{equation}\label{eq:func_eq_ybe}
    W_j(\lambda_{j+1}-\lambda_j)\, D_N(\lambda_1,..,\lambda_j,\lambda_{j+1},..,\lambda_N)=D_N(\lambda_1,..,\lambda_{j+1},\lambda_j,..,\lambda_N)\, W_j(\lambda_{j+1}-\lambda_j)\, .
\end{equation}

\paragraph{} The $N$-site generalized RDM can be restricted to the subspace diagonal in one of the spins, e.g.\ with matrix elements with $\alpha_j=\beta_j$ for the $j$-th index, by inserting the projector $\mathcal{P}_{\text{per}}$ onto states in $\mathcal{H}^L$ obeying \emph{periodic} boundary conditions in (\ref{eq:def_D}) as
\begin{equation}
    \label{eq:def_D-restricted}
    D_N^{(j)}(\lambda_1,\dots,\lambda_{N})^{\{\underline{\bm{\alpha}}\}\{\underline{\bm{\beta}}\}}   
    = 
      \frac{\bra{\Phi} \left(\prod_{k=1}^{j} T^{\alpha_{k-1} \beta_{k-1}}_{\alpha_k \beta_k}(\lambda_k)\right) \mathcal{P}_{\text{per}}\,
      \left(\prod_{k=j+1}^{N}  T^{\alpha_{k-1} \beta_{k-1}}_{\alpha_k \beta_k}(\lambda_k)\right) \ket{\Phi}}
      {{\braket{\Phi}{\Phi}}\,\prod\limits_{k=1}^{N} \Lambda(\lambda_k)}\,.
\end{equation}
Performing partial traces of $D_N^{(1)}$ ($D_N^{(N-1)}$) over the spin $\alpha_0$ ($\alpha_{N}$) one obtains the follwoing relations between RDMs of different order:
\begin{equation}
\label{eq:parttr}
\begin{aligned}
    &\mathrm{tr}_0 \left(D_N^{(1)}(\lambda_1,\dots,\lambda_N)\right) = \sum_{\alpha_0} \delta_{\alpha_0\beta_0}\,\delta_{\alpha_1\beta_1}\, \left[D_{N}(\lambda_1,\dots,\lambda_N)\right]^{\alpha_0\dots\alpha_N,\beta_0\dots\beta_N} = D_{N-1}(\lambda_2,\dots,\lambda_{N})\,,\\
    &\mathrm{tr}_N \left(D_N^{(N-1)}(\lambda_1,\dots,\lambda_N)\right) = \dots = D_{N-1}(\lambda_1,\dots,\lambda_{N-1})\,.
\end{aligned}
\end{equation}
To prove these relations one uses the fact that $\sum_\alpha T_{\gamma\gamma'}^{\alpha\alpha}(u)\, \mathcal{P}_{\text{per}}\,T^{\gamma\gamma'}_{\dots}(v)= \delta_{\gamma\gamma'}\,t(u)\,T^{\gamma\gamma}_{\dots}(v)$.

\paragraph{} For another functional equation we introduce a linear operator $A_N(\lambda_1,\dots,\lambda_N):\mathrm{End}(\mathcal{V}^N)\to\mathrm{End}(\mathcal{V}^N)$ \cite{FrWe21} (cf. \cite{AuKl12} for a similar construction for the six-vertex model):
the action of $A_N$ on an operator $B\in\mathrm{End}(\mathcal{V}^N)$ is
\begin{equation*}
\begin{aligned}
 &\big( A_N(\lambda_1,\dots,\lambda_N)[B]\big)^{\{\underline{\bm{\alpha}}\}\{\underline{\bm{\beta}}\}}
    =\frac{\delta_{\alpha_0\beta_0}\delta_{\alpha_N\beta_N}}{\prod_{j=1}^N \rho(\lambda_j-\lambda_N)\rho(\lambda_N-\lambda_j)}  \times\\[10pt]
&\qquad\scalebox{0.54}{\begin{tikzpicture}[baseline=(current bounding box.center),scale=1]
 
  \def \b {12} 
  \def \c {\b/6} 
  \def \d {0.5 * 1.41421356237 * \c}
  \def \sc {0.7}
  \def \sch {1.25}
  \def \scp {1.5}
  \draw (\c,2*\c)--(\b-2*\c,2*\c)--(\b-2*\c,-2*\c)--(\c,-2*\c)--(\c,2*\c);
  \draw (\c+\b,-1*\c)--(\c-\c+\b,-2*\c)--(\c-2*\c+\b,-1*\c)--(\c-\c+\b,-0*\c)--(\c+\b,-1*\c);
  \draw (\c+\b-\c,-1*\c-\c)--(\c-\c+\b-\c,-2*\c-\c)--(\c-2*\c+\b-\c,-1*\c-\c)--(\c-\c+\b-\c,-0*\c-\c)--(\c+\b-\c,-1*\c-\c);
  \draw (\c+\b-\c+3*\c,-1*\c-\c+3*\c)--(\c-\c+\b-\c+3*\c,-2*\c-\c+3*\c)--(\c-2*\c+\b-\c+3*\c,-1*\c-\c+3*\c)--(\c-\c+\b-\c+3*\c,-0*\c-\c+3*\c)--(\c+\b-\c+3*\c,-1*\c-\c+3*\c);
  \draw (\c,-1*\c)--(\c-\c,-2*\c)--(\c-2*\c,-1*\c)--(\c-\c,-0*\c)--(\c,-1*\c);
  \draw (\c-2*\c,-1*\c+2*\c)--(\c-\c-2*\c,-2*\c+2*\c)--(\c-2*\c-2*\c,-1*\c+2*\c)--(\c-\c-2*\c,-0*\c+2*\c)--(\c-2*\c,-1*\c+2*\c);
  \draw[dotted] (\c,0)--(0,0);
  \draw[dotted] (\c,-2*\c)--(0,-2*\c);
  \draw[dotted] (\c,\c)--(-\c,\c);
  \draw[dotted] (\c,\c+\c)--(-2*\c,\c+\c);
  \draw[dotted] (\b-2*\c,-\c)--(\b-\c,-\c);
  \draw[dotted] (\b-2*\c,-\c+\c)--(\b+0*\c,-\c+\c);
  \draw[dotted] (\b-2*\c,-\c+\c+\c)--(\b+\c,-\c+\c+\c);
  \draw[dotted] (\b-2*\c,-\c+\c+\c+\c)--(\b+\c+\c,-\c+\c+\c+\c);
  \node[rotate=45] at (\b+\c,0) {\dots};
  \node[rotate=135] at (-\c,0) {\dots};
  
  \foreach \x in {-1,0,1,2}{\node[draw,circle,inner sep=2pt,fill] at (\c,\x*\c) {};}
  \foreach \x in {-2,-1,0,1,2}{\node[draw,circle,inner sep=2pt,fill] at (\b-2*\c,\x*\c) {};}
  \node[draw,circle,inner sep=2pt,fill] at (0,0) {};
  \node[draw,circle,inner sep=2pt,fill] at (-\c,\c) {};
  \node[draw,circle,inner sep=2pt,fill] at (-2*\c,2*\c) {};
  \node[draw,circle,inner sep=2pt,fill] at (\b-\c,-\c) {};
  \node[draw,circle,inner sep=2pt,fill] at (\b+0*\c,0) {};
  \node[draw,circle,inner sep=2pt,fill] at (\b+\c,\c) {};
  \node[draw,circle,inner sep=2pt,fill] at (\b+2*\c,2*\c) {};

 \node[scale=\sch,left] at (-3*\c,\c) {$\alpha_0$};
 \node[scale=\sch,left] at (-2*\c,0) {$\alpha_1$};
 \node[scale=\sch,left] at (-\c,-\c) {$\alpha_{N-2}$};
 \node[scale=\sch,left] at (0,-2*\c) {$\alpha_{N-1}$};
 \node[scale=\sch,right] at (\b+\c+\c+\c,\c) {$\beta_{0}$};
  \node[scale=\sch,right] at (\b+\c+\c,0) {$\beta_{1}$};
 \node[scale=\sch,right] at (\b+\c,-\c) {$\beta_{N-2}$};
 \node[scale=\sch,right] at (\b,-2*\c) {$\beta_{N-1}$};
 \node[scale=\sch,below] at (\b-\c,-3*\c) {$\alpha_N=\beta_{N}$};
  \node[scale=2] at (2.5*\c,0) {$B$};
    \node[scale=\scp] at (0,-\c) {$\lambda_N-\lambda_{N-1}$};  
    \node[scale=\scp] at (-2*\c,\c) {$\lambda_N-\lambda_1$};
  \node[scale=\scp] at (\b-\c,-2*\c) {$P_-$};
  \node[scale=\scp] at (\b,-\c) {$\lambda_{N-1}-\lambda_{N}$};
  \node[scale=\scp] at (\b+2*\c,\c) {$\lambda_1-\lambda_N$};
 \end{tikzpicture}}\\
\end{aligned}
\end{equation*}
with the operator $P_-$, related to the Boltzmann weight at the crossing parameter:
\begin{equation}\label{eq:p_minus}
\bra{\alpha_0\alpha_1\alpha_2} P_-\ket{\beta_0 \beta_1 \beta_2} = \begin{tikzpicture}[baseline=(current bounding box.center)]
  \def \d {0.5 * 1.41421356237}
  \draw (0,0)--(\d,\d)--(2*\d,0)--(\d,-\d)--(0,0);
  \node [left] at (0,0) {$\alpha_1$};
  \node [above] at (\d,\d) {$\alpha_0=\beta_0$};
  \node [right] at (2*\d,0) {$\beta_1$};
  \node [below] at (\d,-\d) {$\alpha_2=\beta_2$};
  \node at (\d,0) {$P_-$};
\end{tikzpicture} \equiv \delta_{\alpha_0 \beta_0} \delta_{\alpha_2 \beta_2} \, \sqrt{\frac{g_{\alpha_0} {g_{\alpha_2}}}{g_{\alpha_1} g_{\beta_1}}}\, \begin{tikzpicture}[baseline=(current bounding box.center)]
  \def \d {0.5 * 1.41421356237}
  \draw (0,0)--(\d,\d)--(2*\d,0)--(\d,-\d)--(0,0);
  \node [left] at (0,0) {$\alpha_1$};
  \node [above] at (\d,\d) {$\alpha_0$};
  \node [right] at (2*\d,0) {$\beta_1$};
  \node [below] at (\d,-\d) {$\alpha_2$};
  \node at (\d,0) {$\lambda$};
\end{tikzpicture}.
\end{equation}
Acting with $A_N$ on the $N$-site RDM one obtains a discrete difference equation of reduced qKZ type \cite{FrWe21}:
The density operator $D_N(\lambda_1,\dots,\lambda_N)$ is a solution of the functional equation
 \begin{equation}
 \label{eq:thm2}
  A_N(\lambda_1,\dots,\lambda_N)[D_N(\lambda_1,\dots,\lambda_{N-1},{\lambda_N})] =D_N(\lambda_1,\dots,\lambda_{N-1},{\lambda_N+\lambda})
 \end{equation}
if $\lambda_N$ is equal to one of the inhomogeneities, i.e.\ $\lambda_N\in \{u_k\}_{k=1}^L$.
For the proof one considers the action of $A_N$ on $D_{N+1}(\lambda_1,\dots,\lambda_N,\lambda_N+\lambda)$. Performing partial traces over $\alpha_N=\beta_N$ and $\alpha_{N+1}=\beta_{N+1}$, respectively, and using the YBE, unitarity and initial condition for the Boltzmann weights (\ref{eq:thm2}) is obtained. For the RSOS  models it is straightforward to show that the restriction on $\lambda_N$ can be dropped for matrix elements of the RDM where $\alpha_{N-1}=1,r-1$.

\paragraph{} Based on numerical results for small $r$ and $N$ we have proposed an identity relating the asymptotics of $D_N$ to  $D_{N-1}$ in the topological sectors with $d_q(j)=1$ when $\lambda_N$ is sent to $i\infty$ \cite{FrWe21}:
\begin{equation}\label{eq:asympt}
\begin{aligned}
    \lim_{\lambda_N\to i\infty}  &\left[D_{N}(\lambda_1,\dots,\lambda_N)\right]^{\alpha_0\dots\alpha_N,\beta_0\dots\beta_N} \\
    &= \left[D_{N-1}(\lambda_1,\dots,\lambda_{N-1})\right]^{\alpha_0\dots\alpha_{N-1},\beta_0\dots\beta_{N-1}} \frac{\left[D_1\right]^{\alpha_{N-1}\alpha_N,\beta_{N-1}\beta_N}}{\sum_\alpha \left[D_1\right]^{\alpha_{N-1}\alpha,\beta_{N-1}\alpha}}\,.
\end{aligned}
\end{equation}
(Recall that $D_1$ is independent of the spectral parameter $\lambda_1$, see (\ref{D1}).)

\paragraph{} Finally, using $W_j(\lambda)=e_j$ together with crossing and unitary one can prove that the density operator (\ref{eq:def_D}) satisfies the reduction relation (see also \cite{MoHC20})
\begin{equation}
\label{eq:thm1}
 \begin{aligned}
 &\bra{\underline{\bm{\alpha}}}e_{N-1}\, D_{N}(\lambda_1,\dots,\lambda_{N-2},u,u+\lambda)\ket{\underline{\bm{\beta}}}
 =\frac{\prod_{k=1}^L\rho(u-u_k)\rho(u_k-u)}{\Lambda(u)\Lambda(u+\lambda)}\\
 &\quad \times
 \bra{\alpha_0..\alpha_{N-2}}D_{N-2}(\lambda_1,\dots,\lambda_{N-2})\ket{\beta_0..\beta_{N-2}} \,\bra{\alpha_{N-2}..\alpha_N}e_{N-1}\ket{\beta_{N-2}..\beta_N}\,
,
 \end{aligned}
\end{equation}
for arbitrary $u$.

Assuming that the factorization (\ref{RSOSfac}) holds for general $\lambda=\pi/r$ we can  apply (\ref{eq:thm2}) to the two-site density matrix $D_2$ of the RSOS model.  This gives rise to a discrete difference equation  satisfied by the scalar function $\omega(u,v)$ for a given value of $r$.  Based on explicit results for $r\leq7$ we propose the following functional equation for $\omega(u,v)$ for $v\in \{u_k\}$ and general values of $\lambda=\pi/r$
\begin{equation}
\label{eq:func}
    \omega(u,v+\lambda)
    =\frac{\sin^2\lambda}{\cos2 (u-v)-\cos2 \lambda}
    -\frac{\cos (2 (u-v-\lambda ))-\cos (2 \lambda )}{\cos (2 (u-v))-\cos (2 \lambda )}\, \omega(u,v)\,.
\end{equation}
We have checked (\ref{eq:func}) for values of $r$ up to $12$.
Note that, as a consequence of (\ref{eq:asympt}), the function $\omega(u,v)$ vanishes for ${v\to i\infty}$.

\section{\texorpdfstring{$N$}{N}-site density matrices - the algebraic part}

To begin our analysis of the algebraic part we note that, using the asymptotic relation (\ref{eq:asympt}), the elements of $F_{N;\varnothing,\varnothing}$ for $N\geq2$ can be obtained recursively
\begin{equation}
\label{eq:recursion_A}
\begin{aligned}
    \left[F_{N;\varnothing,\varnothing}\right]^{\alpha_0\dots\alpha_N,\alpha_0\dots\alpha_N}
    &= \left[F_{N-1;\varnothing,\varnothing}\right]^{\alpha_0\dots\alpha_{N-1},\alpha_0\dots\alpha_{N-1}}
    \frac{D_1^{[\alpha_{N-1}\alpha_N]}}{\sum_\alpha D_1^{[\alpha_{N-1} \alpha]}}\\
    &= \frac1{2\cos\lambda}\,\left[F_{N-1;\varnothing,\varnothing}\right]^{\alpha_0\dots\alpha_{N-1},\alpha_0\dots\alpha_{N-1}} \frac{\sin\alpha_N\lambda}{\sin\alpha_{N-1}\lambda}\,,
\end{aligned}
\end{equation}
giving
\begin{align}
\label{mel_aN}
   \left[F_{N;\varnothing,\varnothing}\right]^{\alpha_0\dots\alpha_N,\beta_0\dots\beta_N}  = \frac{2\lambda}{\pi}\, \left(\prod_{j=0}^N\delta_{\alpha_j\beta_j}\right)
      \frac{\sin\alpha_0\lambda\,\sin\alpha_N\lambda}{(2\cos\lambda)^N}\,.
\end{align}
Note that $F_{N;\varnothing,\varnothing}$ is proportional to the identity in each of the $[\alpha_0,\alpha_{N}]$-blocks of the $N$-site density matrix $D_N$.

\subsection{The two-site density matrix} 
The two-site density matrices in the topological sectors with $d_q=1$ can be written as
\begin{align}
 \label{D2}
        D_2(\lambda_1,\lambda_2) = F_{2;\varnothing,\varnothing} + F_{2;(1)(2)}\, \omega(\lambda_1,\lambda_2) \equiv F_{2;\varnothing,\varnothing} \left(\mathbf{1}+B_2\,\omega(\lambda_1,\lambda_2) \right)\,,
\end{align}
with $F_{2;\varnothing,\varnothing}$ given  in (\ref{mel_aN}). Numerically we find that the matrix $F_{2;(1),(2)}=F_{2;\varnothing,\varnothing}\,B_2$ is also constant (i.e.\ independent of the spectral parameters).  Therefore it can be obtained by utilizing (\ref{eq:thm2}): since all matrix elements of $D_2$ are given in terms of the unknown elements of $B_2$ and the single function $\omega(\lambda_1,\lambda_2)$ this equation determines the former.
Based on this process we have computed $B_2$ for $r=4,\dots,12$.
%
In terms of the operators $\mathbf{1}$ and $e_1$ forming the basis of the Temperley-Lieb algebra $\text{TL}_2$ the result can be expressed as 
\begin{equation}
\label{eq:B2_TL}
    B_2=2\left(2\cos\lambda\,e_1-\mathbf{1}\right)\,,
\end{equation}
Explicitely, the non-zero blocks of $B_2$ are
\begin{equation}
\begin{aligned}
    &B_2^{[1,1]} = B_2^{[r-1,r-1]} = 2\,\frac{\sin3\lambda}{\sin\lambda}\,,\\
    &B_2^{[aa]} = \frac2{\sin a\lambda}\,\begin{pmatrix}
      \sin(a-2)\lambda & 
      2\cos\lambda\sqrt{\sin(a-1)\lambda\,\sin(a+1)\lambda} \\ 
      2\cos\lambda\sqrt{\sin(a-1)\lambda\,\sin(a+1)\lambda} & 
      \sin (a+2)\lambda 
    \end{pmatrix}\,,\\
    &B_2^{[a-1,a+1]}= B_2^{[a+1,a-1]}=-2\,,
\end{aligned}
\end{equation}
for $2\leq a \leq r-2$.

\subsection{The three-site reduced density matrix} 

According to (\ref{RSOSfac}) the three-site density matrix factorizes as
\begin{equation}
\label{eq:D3_fac}
\begin{aligned}
    D_3(\lambda_1,\lambda_2,\lambda_3) =& F_{3;\varnothing,\varnothing} \left( 1 + B_{3;(1)(2)}(\lambda_1,\lambda_2,\lambda_3)\, \omega(\lambda_1,\lambda_2)\right.\\
    &\left. + B_{3;(2)(3)}(\lambda_1,\lambda_2,\lambda_3)\, \omega(\lambda_2,\lambda_3) + B_{3;(1)(3)}(\lambda_1,\lambda_2,\lambda_3)\, \omega(\lambda_1,\lambda_3)\right)\,.
\end{aligned}
\end{equation}
Here $F_{3;\varnothing,\varnothing}$ has been obtained in (\ref{mel_aN}) before.
Using the explicit construction (\ref{eq:def_D}) of $D_3$ for a system of size $L=2$ we have analyzed the dependence of the coefficient matrices $B_{3;I,J}$ on the spectral parameters $\lambda_j$ using the algorithm presented in Ref.~\cite{FrWe21}.  As a result they are found to be of the form\footnote{The factorization of $D_{N>2}$ for $r=4$ is not unique due to the identity 
\begin{equation*}
    \sin2\lambda_{12}\, \omega(\lambda_1,\lambda_2)
    +\sin2\lambda_{23}\, \omega(\lambda_2,\lambda_3)
    =\sin2\lambda_{13}\, \omega(\lambda_1,\lambda_3)\,,
\end{equation*}
satisfied by the two-site function $\omega$ \cite{FrWe21}.  This implies that $f^4$ does not enter (\ref{eq:D3_fac}) and therefore may be chosen to be zero.}
\begin{equation}
\label{B3_par}
\begin{aligned}
    B_{3;(1)(2)}(\lambda_1,\lambda_2,\lambda_3) &= f_{12}^1 + \left(\cot\lambda_{13} - \cot\lambda_{23}\right) f^2 + (1+\cot\lambda_{13}\,\cot\lambda_{23})\, f^4\,,\\
    B_{3;(2)(3)}(\lambda_1,\lambda_2,\lambda_3) &= f_{23}^1 + \left(\cot\lambda_{12} - \cot\lambda_{13}\right) f^2 + (1+\cot\lambda_{12}\,\cot\lambda_{13})\, f^4\,,\\
    B_{3;(1)(3)}(\lambda_1,\lambda_2,\lambda_3) &= f_{13}^1 + \left(\cot\lambda_{23} - \cot\lambda_{12}\right) f^2 + (1-\cot\lambda_{23}\,\cot\lambda_{12})\, f^4\,,
\end{aligned}
\end{equation}  
with constant matrices $f^1_{ij}=\left(f^1_{ij}\right)^\top$, $f^2=-\left(f^2\right)^\top$ and $f^4=\left(f^4\right)^\top$.  $f^1_{ij}$ and $f^2$ can be computed using Eq.~(\ref{eq:asympt})
together with the YB relation (\ref{eq:func_eq_ybe}): 
%
since $\cot x\to-i$ and $\omega(y,x)\to0$ in the limit $x\to i\infty$ one can express $f^1_{12}$ in terms of $B_2$:
\begin{equation}
\label{eq_red12}
\begin{aligned}
    \left[f_{12}^1\right]^{\alpha_0\dots\alpha_2\alpha_3,\alpha_0\dots\beta_2\alpha_3} 
    &= \frac1{\left[F_{2;\varnothing,\varnothing}\right]^{\alpha_0\dots\alpha_3,\alpha_0\dots\alpha_3}}\delta_{\alpha_2\beta_2}\left[F_{2;\varnothing,\varnothing}\,B_2\right]^{\alpha_0\dots\alpha_2,\alpha_0\dots\alpha_2}
      \frac{D_1^{[\alpha_{2}\alpha_3]}}{\sum_\alpha D_1^{[\alpha_{2} \alpha]}}\\
    &= \delta_{\alpha_2\beta_2}\left[B_2\right]^{\alpha_0\dots\alpha_2,\alpha_0\dots\alpha_2}
\end{aligned}
\end{equation}
Similar relations determining $f^1_{23}$, $f^1_{13}$, and $f^2$ follow after using the YB-relation (\ref{eq:func_eq_ybe}):
\begin{align*}
  \lim_{\lambda_3\to i\infty}D_3(\lambda_1,\lambda_3,\lambda_2) &= \lim_{\lambda_3\to i\infty}W_2(\lambda_3-\lambda_2)\cdot D_3(\lambda_1,\lambda_2,\lambda_3)\cdot \left[W_2(\lambda_3-\lambda_2)\right]^{-1}\,,\\
  \lim_{\lambda_3\to i\infty}D_3(\lambda_3,\lambda_1,\lambda_2) &= \lim_{\lambda_3\to i\infty}W_1(\lambda_3-\lambda_1)\cdot D_3(\lambda_1,\lambda_3,\lambda_2)\cdot \left[W_1(\lambda_3-\lambda_1)\right]^{-1}\,.
\end{align*}
Note that in the limit we have
\begin{equation*}
    \lim_{u\to i\infty}\mathrm{e}^{iu}\,W_j(u)= \frac{1}{2i\sin\lambda}\left( \mathrm{e}^{-i\lambda}\mathbf{1}-e_j \right)\,,
    \quad
    \lim_{u\to i\infty}\left[\mathrm{e}^{iu}\,W_j(u)\right]^{-1} \to {2i\sin\lambda}\left( \mathrm{e}^{i\lambda}\mathbf{1}-e_j \right)\,,
\end{equation*}
where $e_j$ is the Temperley-Lieb operator acting on the spins $\alpha_{j-1},\alpha_j\alpha_{j+1}$.  
The remaining coefficients in $f^4$ are determined by the reduced qKZ equation (\ref{eq:thm2}) for $D_3$.
Collecting the results for $r$ up to $9$ we find that, similar as for $D_2$ above, $B_{3;I,J}$ can be expanded in the basis $\{\mathbf{1}, e_{1}, e_2, e_1e_2, e_2e_1\}$ of the Temperley-Lieb algebra $\text{TL}_3$ :
\begin{equation}
\label{eq:B3_TL}
\begin{aligned}
    f^1_{12} &=2\left(2\cos\lambda\,e_1 -\mathbf{1}\right)\,,\\
    f^1_{23} &=2\left(2\cos\lambda\,e_2 -\mathbf{1}\right)\,,\\
    f^1_{13} &=2\left(2\cos\lambda\,\left(e_1+e_2\right) -2\cos^2\lambda\,\left(e_1e_2+e_2e_1\right)-\mathbf{1}\right)\,, \\
    f^2 &= \sin2\lambda\,\left(e_1e_2-e_2e_1\right)\,,\\
    f^4 &= 2\,\frac{1-\cos^2\lambda}{1-2\cos^2\lambda}\,\left(\mathbf{1} - 2\cos\lambda\,(e_1+e_2) + 2\cos^2\lambda\,(e_1e_2+e_2e_1) \right)\,.
\end{aligned}
\end{equation}

\subsection{The four-site reduced density matrix} 
Similar as in (\ref{XXXfac}) for the Heisenberg model we expect that the $4$-site reduced density matrix $D_4$ factorizes as (see also \cite{BoST05,SaST05})
\begin{equation}
\label{d4fac}
\begin{aligned}
   &D_4(\lambda_1,\dots,\lambda_4) = F_{4;\varnothing,\varnothing}\\
   &\quad + F_{4;(1)(2)}(\{\lambda_j\})\,\omega(\lambda_1,\lambda_2) + F_{4;(1)(3)}(\{\lambda_j\})\,\omega(\lambda_1,\lambda_3) + F_{4;(1)(4)}(\{\lambda_j\})\,\omega(\lambda_1,\lambda_4)\\
   &\quad + F_{4;(2)(3)}(\{\lambda_j\})\,\omega(\lambda_2,\lambda_3) + F_{4;(2)(4)}(\{\lambda_j\})\,\omega(\lambda_2,\lambda_4) + F_{4;(3)(4)}(\{\lambda_j\})\,\omega(\lambda_3,\lambda_4)\\
   &\quad + F_{4;(1,2)(3,4)}(\{\lambda_j\})\,\omega(\lambda_1,\lambda_3)\,\omega(\lambda_2,\lambda_4) + F_{4;(1,2)(4,3)}(\{\lambda_j\})\,\omega(\lambda_1,\lambda_4)\,\omega(\lambda_2,\lambda_3)\\
   &\quad + F_{4;(1,3)(2,4)}(\{\lambda_j\})\,\omega(\lambda_1,\lambda_2)\,\omega(\lambda_3,\lambda_4)\,,
\end{aligned}
\end{equation}
with the nearest neighbour two-site function $\omega(\lambda_k,\lambda_\ell)$ introduced in Eq.~(\ref{RSOSfac}) and $F_{4;\varnothing,\varnothing}$ given in (\ref{mel_aN}).  To verify such a decomposition we have computed the structure functions $F_{4;I,J}$ for a single matrix element of $D_4$ in the $r=5$ RSOS model with $L=4$ sites and found that they are elementary functions of the differences $\lambda_{k\ell}=\lambda_k-\lambda_\ell$.  For an expansion of the structure functions in the basis of the Temperley-Lieb algebra $\text{TL}_4$ (similar as for $D_2$ and $D_3$) one needs these data for all matrix elements in blocks $[a,a]$, $[a,a\pm2]$ of $D_4$ as input.  We leave this to a future publication.

\section{The nearest neighbour function \texorpdfstring{${\omega(u,v)}$}{omega(u,v)}} 
It remains to determine the nearest neighbour function $\omega(u,v)$ which depends on the model specific parameters, i.e.\ system size and inhomogeneities $u_k$, and the state of the system.  From its definition (\ref{D2}) in terms of the two-site density matrix and our observations above $\omega$ has the following properties:
\begin{enumerate}
    \item[(a)] periodicity: $\omega(u,v)=\omega(u+\pi,v)=\omega(u,v+\pi)$,
    \item[(b)] asymptotics: $\lim_{v\to i\infty} \omega(u,v)=0$,
    \item[(c)] analyticity: as a consequence of (\ref{eq:def_D}) poles of $\omega(u,v)$ correspond to zeroes of the transfer matrix eigenvalues $\Lambda(u)$, $\Lambda(v)$.
\end{enumerate}
Restricting ourselves to the ground state of the quantum RSOS model (\ref{eq:HqRSOS}) our numerical studies of the finite-size expressions show that $\omega(u,v)$ is an analytical function of both $u$ and $v$ in the strips $\mathcal{S}_0=\left\{z\in\mathbb{C}:-\lambda/2\lesssim \mathrm{Re}(z)\lesssim 3\lambda/2\right\}$ (the physical strip for the RSOS models in regime III/IV) and $\mathcal{S}_1 = \left\{z\in\mathbb{C}: -\pi+3\lambda/2\lesssim \mathrm{Re}(z) \lesssim -\lambda/2\right\}$. Within these strips we find that $\omega(u,v)$ depends on $x=u-v$ only when $L\to\infty$. 

With these data as input we can now solve the functional equation (\ref{eq:func}) for $\omega(u,v)$ derived in Section~\ref{sec:funceq}.\footnote{It is straightforward to extend this procedure to other eigenstates of the RSOS transfer matrix in this topological sector by taking into account the additional poles of $\omega(u,v)$ in the strips $\mathcal{S}_{0,1}$.}
 
The RSOS model for $r=4$ can be mapped to the Ising model.  In this case we have an explicit expression for the product $\Lambda(u)\Lambda(u+\lambda)$ \cite{ObPW96}
\begin{equation}
  \Lambda(u)\Lambda(u+\lambda) = \prod_{k=1}^L\rho(u-u_k)\rho(u_k-u) +y \prod_{k=1}^L\rho(u+\lambda-u_k)\rho(u_k-u-\lambda)
\end{equation}
where $y=\pm1$ is the eigenvalue of the height reflection operator mapping heights $a\to r-a$.  In the ground state of the RSOS hamiltonian $y=(-1)^{L/2}$.  Therefore we can use the reduction relation (\ref{eq:thm1}) to compute the function $\omega_{r=4}(u,v)$: using the factorized form (\ref{eq:D3_fac}) of the three-site density matrix together with the explicit form $D_1(u)=\frac14\mathbf{1}$ (for $r=4$) we obtain the difference equation
\begin{equation}
\label{eq:func4}
    \sin2(u-v)\,\omega_4(u,v) + \cos2(u-v)\,\omega_4(u,v+\lambda) = r_L(v) \equiv \frac{1}{\left(1+y\,\tan^{L}2v\right)}-\frac12\,,
\end{equation}
for the homogeneous model, $u_k\equiv0$.  Note that, unlike the reduced qKZ equation (\ref{eq:func}), this equation holds for arbitrary values of $u$ and $v$. In addition we find an explicit expression for
\begin{equation}\label{eq:funcr4_explicit}
    \omega_4(u,u+\lambda) = r_L(u)\,.
\end{equation}
This allows for a computation of the nearest neighbour functions for \textit{arbitrary finite} chain lengths.  
To this end, we need to solve (\ref{eq:func4}).  We consider $y=+1$ for the ground state of the  RSOS model with $L\in4\mathbb{N}$ and define $h(u,v)\equiv \sin(2(u-v)) \omega(u,v)$. Setting $a(u,z)\equiv h\left(u,iz+{\pi}/{8}\right)$ we arrive at the difference equation
\begin{equation}\label{eq:func4_rewritten}
    a\left(u,z+i\frac{\pi}{8}\right) - a\left(u,z-i\frac{\pi}{8}\right) = r_L(iz)\,. 
\end{equation}
Numerical studies of small system sizes reveal that $a$ has non-zero asymptotics, $\lim_{z\to \infty} a(u,z) \neq0$. Hence, we take the derivative of (\ref{eq:func4_rewritten}) and use Fourier methods to obtain $a(u,z)$ up to a $u$-dependent term
\begin{equation}
    a(u,z)=-\int\limits_{-\infty}^\infty r_L(iy)\, 
    K(z-y)\, \mathrm{d}y + \Psi(u)\, ,
\end{equation}
with a kernel given by
\begin{equation}
    K(z)=\int\limits_{-\infty+i0}^{\infty+i0} \frac{e^{ikz}}{4\pi \sinh\left(\frac{\pi k}{8}\right)}\, \mathrm{d}k\, .
\end{equation}
Using (\ref{eq:funcr4_explicit}) we can determine $\Psi(u)$ and finally find
\begin{equation}\label{eq:solutionr4}
    h(u,v)=\frac{2}{\pi} \int\limits_{-\infty}^\infty \frac{r_L(iy)\,
    \sin\left(4(v-u)\right)}{\sinh\left(4(y+i u)\right)\sinh\left(4(y+i v)\right)}\, \mathrm{d} y
\end{equation}
for $u,v\in(0,\pi/4)$ (outside of these intervals $h(u,v)$ is obtained by analytical continuation). 
Note that the chain length $L$ only enters (\ref{eq:solutionr4}) as a parameter in the function $r_L(iy)$. 
In the thermodynamic limit $r_\infty(iy)\equiv +1/2$ for $|\text{Im}(y)|<\pi/8$ and we find
\begin{equation}
\label{eq:ff4}
\begin{aligned}
    \omega_4(u,v) &= \frac{2(u-v)-\pi(k-\ell)}{\pi\sin(u-v)}\qquad
    \text{for~}u-\frac{k\pi}2\,,\, v-\frac{\ell\pi}2 \in\left(-\frac{\pi}{8},\frac{3\pi}{8}\right)
   \,,\quad k,\ell\in\mathbb{Z}\,.
\end{aligned}
\end{equation}

\begin{figure}[t]
    \centering
    \includegraphics[width=0.8\textwidth]{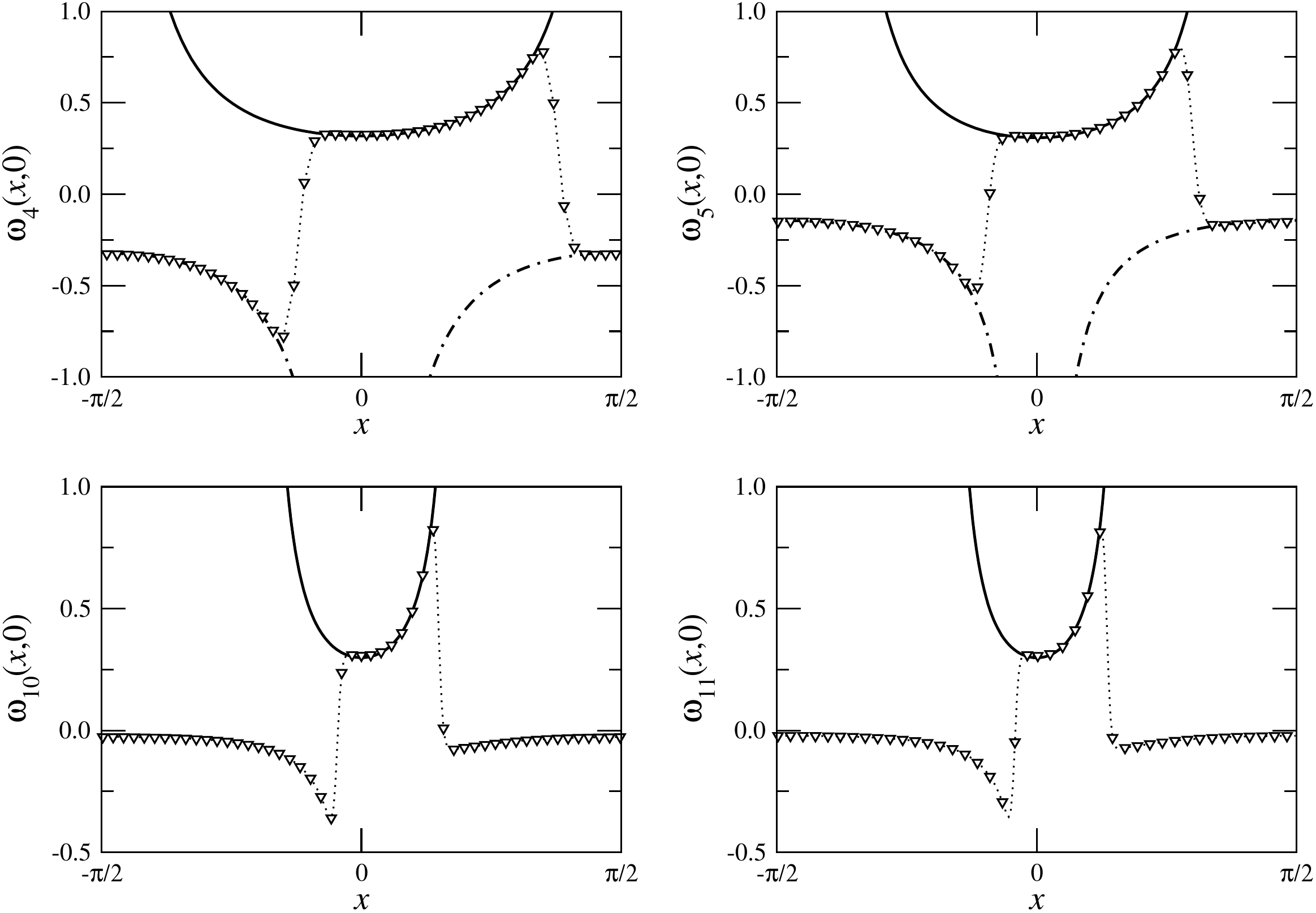}
    \caption{The two-site function $\omega_r(x,0)$ for the ground state of the quantum RSOS model (\ref{eq:HqRSOS}) with $r=4,5,10,11$ in the thermodynamic limit. Solid lines are solutions based on the conjectures (\ref{eq:ff_conj}) in the thermodynamic limit, dotted lines and symbols ($\triangledown$) are numerical values for $L=8$.  Broken lines are the solutions $\widetilde{\omega}_{r,k}(x)$ (\ref{eq:ff_other}) to the functional equation (\ref{eq:func}) for $r=4,5$ in the thermodynamic limit.}
    \label{fig:ff}
\end{figure}

For $r\geq5$ the two-site function $\omega$ can be obtained from the explicit form of $D_2$ in terms of the operators $T^{\alpha\beta}_{\gamma\delta}$ (for small systems) or by solving the functional equation (\ref{eq:func}).
Assuming that (\ref{eq:func}) holds throughout the analyticity strips $\mathcal{S}_0$ and $\mathcal{S}_1$ in the thermodynamic limit one can solve the functional equation using Fourier methods. Based on our results for small $r$ for $x\in\mathcal{S}_0$ we conjecture
\begin{equation}
\label{eq:ff_conj}
    \omega_r(x) = \frac{\cos2x-\cos2\lambda}{\sin rx}
    \begin{cases}
      \left(\frac{\cos\lambda}{\lambda\,\sin3\lambda}\, x+\sum_{j=1}^{(r-4)/2} a_j\,\sin 2jx\right) & \quad \text{for~}r\geq4\text{~even}\\
      \left(\sum_{j=0}^{(r-5)/2} a_j\,\sin (2j+1)x\right)  & \quad \text{for~}r\geq5\text{~odd}
      \end{cases}\,.
\end{equation}
Note that this branch of the two-site function is continuous in the interval $-2\lambda<x<2\lambda$.
Plugging this expression into (\ref{eq:func}) it is straightforward to solve for the coefficients $a_j$, see Table~\ref{tab:ff} for $r\leq10$.
\begin{table}
\caption{\label{tab:ff}Coefficents $a_j$ in the conjectures for the two-site function $f$ (\ref{eq:ff_conj}).}
\begin{ruledtabular}
\begin{tabular}{cl}
$r$ & $a_j$ \\\hline
$5$ & $a_0 = \sqrt{5}$ \\ 
$6$ & $a_1=1$ \\
$7$ & $a_0 = 2\cos(\pi/7)+2$,
      $a_1 = 4\cos(\pi/7)-3 $ \\ 
$8$ & $a_1 = 2$, $a_2 = \sqrt{2}-1$ \\
$9$ & $a_0 = (10\cos(\pi/9)+7)/(4\cos(\pi/9)-1)$,
      $a_1 = 2(\cos(\pi/9)+1)/(2\cos(\pi/9)+1)$,\\
    & $a_2 = (2\cos(\pi/9)-1)/(2\cos(\pi/9)+1)$\\
$10$ & $a_1 = \sqrt{5}+1$,
       $a_2 = 1$,
       $a_3 = \sqrt{5}-2$
\end{tabular}
\end{ruledtabular}
\end{table}
Comparing (\ref{eq:ff_conj}) with finite size data for the two-site function one finds rapid convergence in the interval $\lambda/2<x<3\lambda/2$, see Fig.~\ref{fig:ff}. At the boundaries of this interval we observe a transition of $\omega$ (smooth for $L\mod4=0$, singular for $L\mod4=2$) to different solutions of the functional equation.  For $r=4$ we have explicit expressions $\widetilde{\omega}_{4,k}(x)$ for these solutions in the thermodynamic limit from Eq.~(\ref{eq:ff4}) for $r=4$ in the intervals around $x-k\pi\in \mathcal{S}_1$, i.e.\
\begin{subequations}
\label{eq:ff_other}
\begin{equation}
    \widetilde{\omega}_{4,k}(x) = \frac{2x-(2k-1)\pi}{\pi\sin2x}\,,
\end{equation}
Solving the functional equation (\ref{eq:func}) for $r=5$ for $x-k\pi\in\mathcal{S}_1$ we obtain:
\begin{equation}
    \widetilde{\omega}_{5,k}(x) = \sqrt{5}\, \frac{\left(\cos2x-\cos2\lambda\right)\left(\sin  x +(-1)^k\sin2\lambda\right)}{\sin5x}\,.
\end{equation}
\end{subequations}
Note that the functions $\widetilde{\omega}_{r,k}(x)$ are regular in the intervals $(k-1)\pi<x<k\pi$.

\section{Multi-point local-height probabilities}
Given the "factorized" expressions (\ref{D1}), (\ref{D2}), (\ref{eq:D3_fac}), \dots\ one obtains the physical correlation functions of the (homogeneous) RSOS models in the limit $\lambda_j\to0$, $j=1,\dots,N$.  In that limit the elements of the density matrices can be expressed in terms of the nearest neighbour correlation function $\omega(u,v)$ and its derivatives at $(u,v)=(0,0)$ which we denote as $\omega_{k,\ell}\equiv\partial_u^k\,\partial_v^\ell\,\omega(u,v)\vert_{u=v=0}$ in the following.

The single-site density matrix in the topological sectors with $d_q=1$ is diagonal and independent of the spectral parameters, hence the two-point LHP for heights $a$, $a+1$ on adjacent sites is given by the matrix elements of $D_1$ (\ref{D1}). 

Similarly, the two-site density matrix in these sectors depends on the spectral parameters through the nearest-neighbour function $\omega(u,v)$ only.  Therefore the three-point LHP for heights $a,b,c$ on three neighbouring sites is given by the diagonal elements of $D_2(0,0)$, (\ref{D2}) with Eqs.~(\ref{mel_aN}) and (\ref{eq:B2_TL}), i.e.
\begin{equation}
\label{LHP3}
    P_{a,b,c} = \bra{a,b,c} D_2(0,0) \ket{a,b,c}
    = \frac{2\lambda}{\pi}\,\frac{\sin a\lambda\,\sin c\lambda}{4\cos^2\lambda} \left(1 +  \bra{ a,b,c} B_2 \ket{a,b,c}\, \omega_{0,0}\right)\,,
\end{equation}
if $|a-b|=|b-c|=1$. 
Note that apart from Eq.~(\ref{eq:ff_conj}) $\omega_{0,0}$ can also be determined from the eigenvalues of the one-dimensional quantum RSOS model (\ref{eq:HqRSOS}).
The ground state energy density of this model in the thermodynamic limit is known to be \cite{BaRe89}
\begin{equation}
    \epsilon_0 = \frac{\lambda}{2\pi}\cot\lambda 
                -\frac{1}{2\pi}\int_{-\infty}^{\infty} \mathrm{d}x\,
                                 \frac{\sinh x\,\sinh(r-3)x}{\sinh2x\,\sinh rx}\,.
\end{equation}
On the other hand, the energy of a given state in terms of the corresponding two-site density matrix is
\begin{equation}
    \frac{E}{L} = \frac{\lambda}{4\pi\sin\lambda}\mathrm{trace}\left(D_2(0,0)\,e_j\right)
                = \frac{\lambda}{4\pi\sin2\lambda}\left( 1 +
                2\,\frac{\sin3\lambda}{\sin\lambda}\, \omega_{0,0}\right)\,.
\end{equation}
We have verified that these expressions coincide for the nearest-neighbour functions for the ground state in the thermodynamic limit (\ref{eq:ff_conj}).
In principle, an additional check of (\ref{LHP3}) would be possible by comparison with the trigonometric limit of the corresponding expression for the massive regimes III and IV of the RSOS model from \cite{LuPu96} which are given in terms of a two-fold integral.  While we have not tried this it might also lead to insights on how to factorize the multiple integral representations of $n>3$-point LHPs.

\begin{figure}[t]
    \centering
    \includegraphics[width=0.45\textwidth]{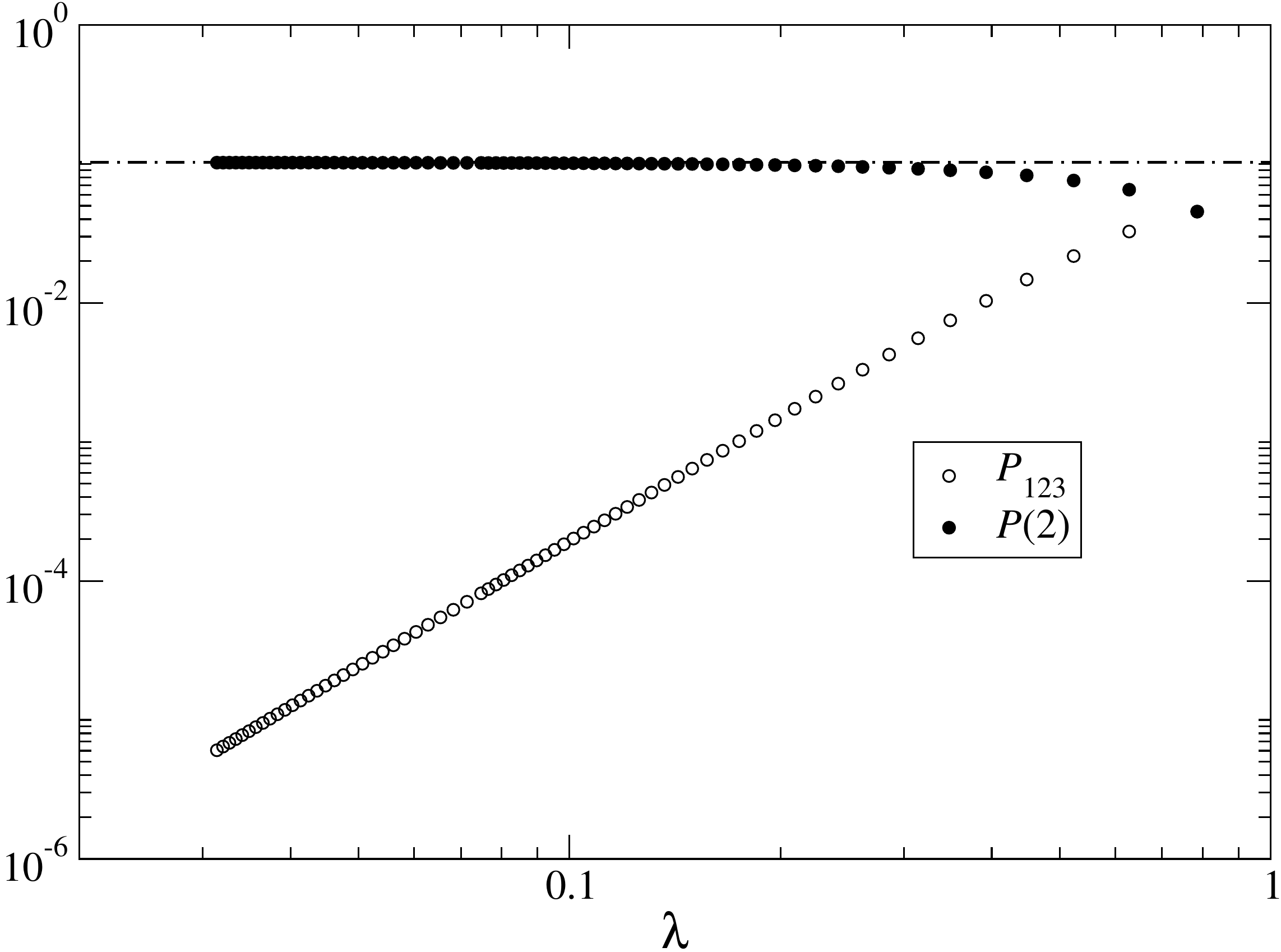}
    \includegraphics[width=0.45\textwidth]{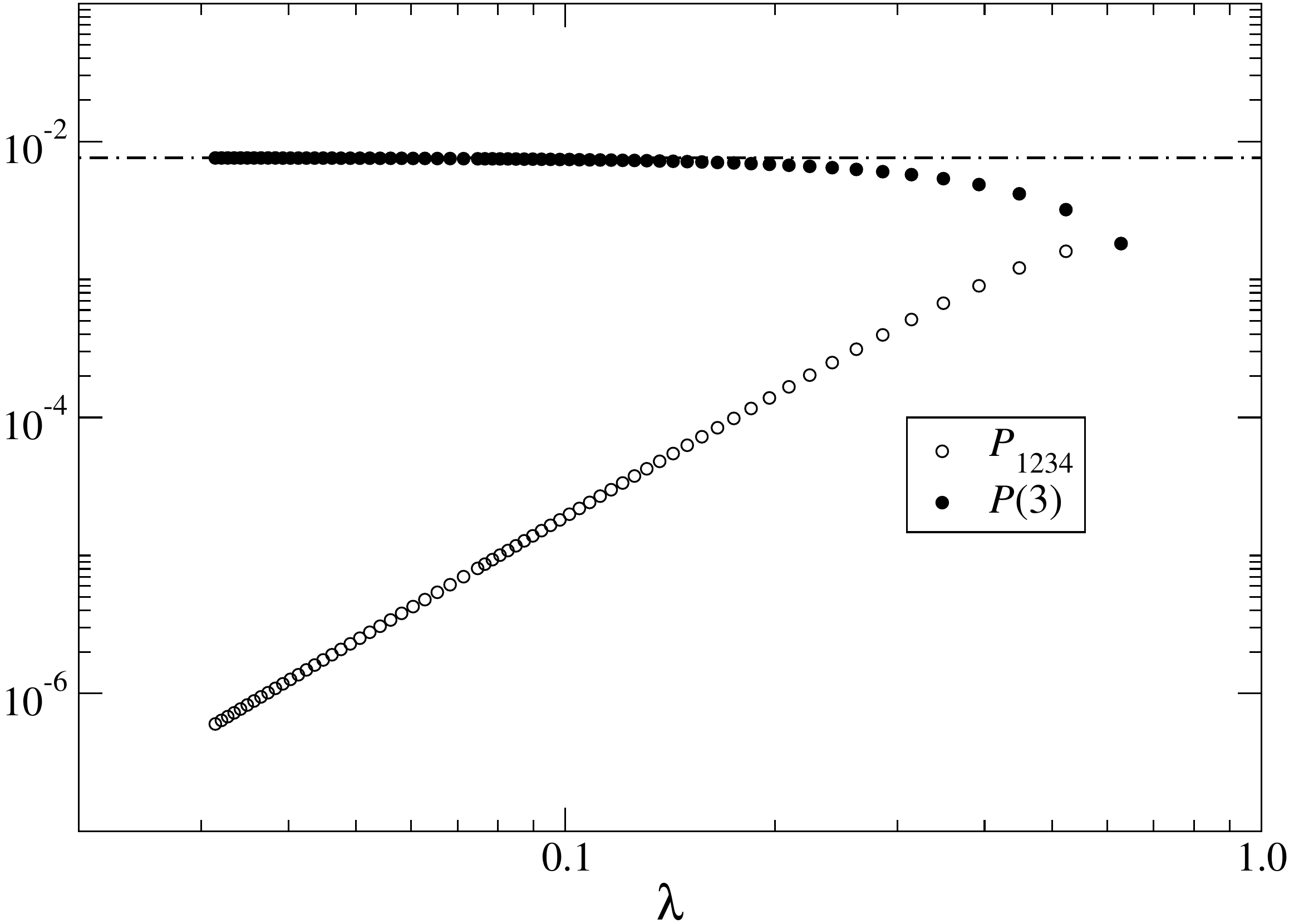}
    
    \caption{Probabilities for the presence of a string of increasing heights in the ground state of the RSOS model (\ref{eq:HqRSOS})  as a function of $\lambda=\pi/r$: displayed are $P_{123}$ and $P(2)=\sum_a P_{a,a+1,a+2}$ (left) and $P_{1234}$ and $P(3)=\sum_a P_{a,a+1,a+2,a+3}$ (right).  For $\lambda\to0$ the $P(n)$ approach the emptiness formation probabilities in the antiferromagnetic Heisenberg XXX chain (\ref{EFP_XXX}, indicated by the dash-dotted lines.}
    \label{fig:p1234}
\end{figure}

Particularly simple multi-point LHPs are obtained from the diagonal elements of $D_N$ in the states $\ket{a,a+1,\dots,a+N}$, i.e.\ the probability for the presence of a segment of length $N$ where the heights are increasing from $a$ to $a+N$: in this case there is no contribution from the Temperley-Lieb operators appearing in the structure functions (\ref{eq:B2_TL}) or (\ref{eq:B3_TL}).  Performing the limit $\lambda_j\to0$ one obtains
\begin{equation}
\begin{aligned}
    &P_{a,a+1,a+2} = \frac{2\lambda}{\pi}\,\frac{\sin a\lambda\,\sin(a+2)\lambda}{4\cos^2\lambda} \left(1-2\,\omega_{0,0}\right)\,,\\
    &P_{a,a+1,a+2,a+3} = \frac{2\lambda}{\pi}\,\frac{\sin a\lambda\,\sin(a+3)\lambda}{8\cos^3\lambda} \Bigg(1 -\frac1{1-2\cos^2\lambda}\Big( 2(1-4\cos^2\lambda)\,\omega_{0,0}\\ &\qquad\qquad\qquad\qquad\qquad\qquad\qquad\qquad\qquad\qquad+\sin^2\lambda\,\left(2\,\omega_{1,1} - \omega_{2,0}\right)\Big)\Bigg)\,.
\end{aligned}
\end{equation}
Using (\ref{eq:ff_conj}) with the coefficients given in Table~\ref{tab:ff} one obtains analytical expressions for these multi-point LHPs for $r$ up to $10$.
%
In Figure~\ref{fig:p1234} the $\lambda=\pi/r$-dependence of $P_{123}$ and $P_{1234}$ is shown for $r$ up to $100$ based on the numerical solution of the reduced qKZ equation (\ref{eq:func}. 
While these quantities vanish as a power law as $\lambda=\pi/r\to0$ the probability for any increasing sequence of heights in the corresponding state, obtained by summation of $P_{a,a+1,\dots,a+n}$ over $a$:
\begin{equation}
    P(n) \equiv \sum_{a=1}^{r-n-1} P_{a,a+1,\dots,a+n}
         = \frac1{4\sin^2\lambda}\left(2+(r-n-1)\left(1-\frac{\sin(n-1)\lambda} {\sin(n+1)\lambda}\right)\right)P_{1,2,\dots,n+1}\,,
\end{equation}
approaches a finite value in this limit.
Note that this quantity corresponds to the emptiness formation probability (EFP$_n$) of finding $n$ adjacent spins up in the vacuum state of the dynamical vertex model corresponding to the RSOS model \cite{KIEU94}. As mentioned above the quantum RSOS model (\ref{eq:HqRSOS}) has also been interpreted as a ferromagnetic chain of interacting non-Abelian $su(2)_k$ anyons with spin $1/2$ for $k=r-2$ \cite{FTLT07,GATL09,GATH13}.  In this context $P(n)$ is the probability that $n\leq k$ neighbouring spin-$1/2$ anyons fuse into a spin-$n/2$ anyon.

With (\ref{D1}) one obtains $P(1)\equiv\frac12$ as expected from the symmetry of the RSOS model under height reflection $a\to r-a$. Surprisingly, there appears a relation between the \emph{ferro}magnetic model considered here and the isotropic spin-$1/2$ Heisenberg \emph{antiferro}magnet: 
we find that the limiting values of $P(2)$ and $P(3)$ for $\lambda\to0$ are the EFP for the latter \cite{BoKo01}, see Fig.~\ref{fig:p1234}:
\begin{equation}
\label{EFP_XXX}
    \begin{aligned}
    P(2) &\to \text{EFP}_2= \frac13-\frac13\ln2 \approx 0.102284273\,,\\
    P(3) &\to \text{EFP}_3= \frac14-\ln2+\frac38\,\zeta(3) \approx 0.007624158\,.
    \end{aligned}
\end{equation}
Here $\zeta(s)=\sum_{n=1}^\infty n^{-s}$ is the Riemann zeta function.

\section{Conclusion}
We have studied reduced density matrices of the critical RSOS models for segments of up to four adjacent sites in finite and infinite chains based on their factorization in terms of a nearest-neighbour two-point function and a set of structure functions.  The latter are independent of the states in the topological sector with quantum dimension $d_q=1$ of the models and can be expressed in terms of the generators of the underlying Temperley-Lieb algebra with prefactors depending on the representation.  For the ground state of the quantum RSOS model (\ref{eq:HqRSOS}) in the thermodynamic limit an explicit expression for the two-point function has been obtained which solves the reduced qKZ equation (\ref{eq:func}).  As an application of our results we have obtained compact expressions for several multi-point local height probabilities in this state.

An essential prerequisite for this work -- apart from the construction of the functional equation (\ref{eq:thm2}) for the $N$-site density matrices in general face models \cite{FrWe21} -- has been a suitable ansatz for their factorization: here we have concentrated ourselves on states from the $d_q=1$ topological sector of the RSOS model where the factorized expressions were of a similar form as (\ref{XXXfac}) known from the isotropic Heisenberg model. For the other sectors our previous work \cite{FrWe21} strongly indicates that the main difference is that the physical part of the RDMs is described in terms of two nearest neighbour functions rather than the single function $\omega$ -- similar as for the XXZ spin chain.  It appears to be worthwhile to study this along the lines used here, i.e.\ assuming that the algebraic part can again be expanded in terms of Temperley-Lieb operators.  Going beyond the critical phases of the RSOS models a further step towards a better understanding the role of integrable structures for correlation functions in face models is the identification of the factorization of the RDMs for the massive regimes. This will provide an alternative to the description of multi-point local height probabilities in terms of multiple integrals \cite{LuPu96}.

\begin{acknowledgments}
The authors thank Alexi Morin Duchesne and Frank Göhmann for valuable discussions. This work is part of the programme of the research unit Correlations in Integrable Quantum Many-Body Systems (FOR 2316).
\end{acknowledgments}

\appendix

\section{The generalized reduced density matrix}
\label{app:DN_graphical}
Using the graphical notation introduced in Eqs.~(\ref{eq:rsos_weights}) and (\ref{eq:def_monodromy}) the generalized reduced density matrix (\ref{eq:def_D}) can be depicted as:
\begin{equation}
 \begin{aligned}
     D_N(\lambda_{1},\dots,\lambda_{N})^{\{\underline{\bm{\alpha}}\}\{\underline{\bm{\beta}}\}} &=
     \frac{\bra{\Phi} \prod_{k=1}^{N}  T^{\alpha_{k-1} \beta_{k-1}}_{\alpha_k \beta_k}(\lambda_k) \ket{\Phi}}
      {{\braket{\Phi}{\Phi}}\,\prod\limits_{k=1}^{N} \Lambda(\lambda_k)}\\
 &=\begin{tikzpicture}[baseline=(current bounding box.center)]
  \def \b {5} 
  \def \c {3} 
  \def \n {\b/3} 
 \draw (0,0)--(\b,0);
 \draw (0,-\c/2)--(\b,-\c/2);
  \draw (0,\c/2)--(\b,\c/2);
   \foreach[evaluate=\x as \yeval using \n*\x] \x in {0,1,2,3}
  \draw (\yeval,-\c/1)--(\yeval,\c/2);
  \node [left] at (0,\c/2) {$\alpha_{0}$};
    \node [right] at (\b,\c/2) {$\beta_{0}$};
 \node [left] at (0,0) {$\vdots$};
 \node [left] at (0,-\c/2) {$\vdots$};
  \node [left] at (0,-\c/1) {$\alpha_{N}$};
  \node [right] at (\b,0) {$\vdots$};
 \node [right] at (\b,-\c/2) {$\vdots$};
  \node [right] at (\b,-\c/1) {$\beta_{N}$};
 \draw (0,\c/2)--(\b,\c/2)--(\b/2,\c/2+\c/4)--(0,\c/2);
   \draw (0,-\c/1)--(\b,-\c/1)--(\b/2,-\c/1-\c/4)--(0,-\c/1);
  \node at (\b/2,\c/2+\c/8) {$\Phi$};
  \node at (\b/2,-\c/1-\c/8) {$\Phi$};

    \node  at (\n/2,\c/4) {\scriptsize $\lambda_{1}-u_1$};
    \node  at (\n/2+\n,\c/4) {\dots};
     \node  at (\n/2+\n,-\c/4) {\dots};
     \node  at (\n/2+\n,-9/4) {\dots};
     \node  at (\n/2+2*\n,\c/4) {\scriptsize $\lambda_{1}-u_L$};
     \node  at (\n/2+2*\n,-\c/4) {\scriptsize $\dots$};
    \node  at (\n/2,-\c/4) {\scriptsize$\dots$};
    \node at (\n/2,-9/4){\scriptsize $\lambda_{N}-u_1$};
     \node  at (\n/2+2*\n,-9/4) {\scriptsize $\lambda_{N}-u_L$};
    \node[draw,circle,inner sep=1pt,fill] at (\n,0) {};
    \node[draw,circle,inner sep=1pt,fill] at (2*\n,0) {};
     \node[draw,circle,inner sep=1pt,fill] at (\n,-3/2) {};
      \node[draw,circle,inner sep=1pt,fill] at (2*\n,-3/2) {};
 
    \end{tikzpicture}\cdot \frac{1}{\braket{\Phi}{\Phi}\prod\limits_{k=1}^{N} \Lambda(\lambda_k)}
 \end{aligned}\,
\end{equation}
where the projection onto the eigenstate $\ket{\Phi}$ of the transfer matrix is indicated by sandwiching of $\prod_{k=1}^N T_{\alpha_k\beta_k}^{\alpha_{k-1}\beta_{k-1}}$.

\vspace*{\fill}

\bibliographystyle{apsrev4-1}

%
\end{document}